%% file: main_arxiv.tex
\documentclass[default]{sn-jnl}

\input{packages}
\input{macros}
\usepackage{enumitem}
\usepackage{amsfonts}
\setlist[itemize,1]{label=$\bullet$}

\begin{document}

\title[Phase retrieval via Diffusion model]{Generalizable Holographic Reconstruction via Amplitude-Only Diffusion Priors}


\author[1]{\fnm{Jeongsol} \sur{Kim}}
\equalcont{These authors contributed equally to this work.}

\author[1]{\fnm{Chanseok} \sur{Lee}}
\equalcont{These authors contributed equally to this work.}

\author[2]{\fnm{Jongin} \sur{You}}

\author*[3]{\fnm{Jong Chul} \sur{Ye}}\email{jong.ye@kaist.ac.kr}

\author*[1,4]{\fnm{Mooseok} \sur{Jang}}\email{mooseok@kaist.ac.kr}

\affil[1]{\orgdiv{Department of Bio and Brain Engineering}, \orgname{KAIST}, \orgaddress{\street{291, Daehakro}, \city{Yuseong-gu}, \postcode{34141}, \state{Daejeon}, \country{South Korea}}}

\affil[2]{\orgdiv{Department of Mechanical Engineering}, \orgname{KAIST}, \orgaddress{\street{291, Daehakro}, \city{Yuseong-gu}, \postcode{34141}, \state{Daejeon}, \country{South Korea}}}

\affil[3]{\orgdiv{Kim Jaechul Graduate School of AI}, \orgname{KAIST}, \orgaddress{\street{291, Daehakro}, \city{Yuseong-gu},\postcode{34141}, \state{Daejeon},  \country{South Korea}}}

\affil[4]{\orgdiv{KAIST Institute for Health Science and Technology}, \orgname{KAIST}, \orgaddress{\street{291, Daehakro}, \city{Yuseong-gu}, \postcode{34141}, \state{Daejeon}, \country{South Korea}}}


\abstract{Phase retrieval in inline holography is a fundamental yet ill-posed inverse problem due to the nonlinear coupling between amplitude and phase in coherent imaging. We present a novel off-the-shelf solution that leverages a diffusion model trained solely on object amplitude to recover both amplitude and phase from diffraction intensities. Using a predictor–corrector sampling framework with separate likelihood gradients for amplitude and phase, our method enables complex field reconstruction without requiring ground-truth phase data for training. We validate the proposed approach through extensive simulations and experiments, demonstrating robust generalization across diverse object shapes, imaging system configurations, and modalities, including lensless setups. Notably, a diffusion prior trained on simple amplitude data (e.g., polystyrene beads) successfully reconstructs complex biological tissue structures, highlighting the method’s adaptability. This framework provides a cost-effective, generalizable solution for nonlinear inverse problems in computational imaging, and establishes a foundation for broader coherent imaging applications beyond holography.}



\maketitle

\section{Introduction}\label{sec1}

Computational imaging \cite{mait2018computational, brady2009optical} redefines conventional image formation by intentionally breaking the direct one-to-one mapping between object and sensor. Instead of passively capturing light intensity, it encodes additional physical information—such as phase, depth, or polarization—through engineered optics, illumination, and sensor design, and decodes it using algorithmic reconstruction. Techniques such as coded-aperture imaging \cite{
levin2007image, arce2013compressive, fenimore1978coded}, single-pixel cameras \cite{edgar2019principles, duarte2008single, sun20133d}, non-line-of-sight imaging \cite{faccio2020non, liu2019non}, lensless holography \cite{ozcan2016lensless, wu2018lensless}, and imaging through scattering media \cite{katz2014non, choi2012scanner, kwon2025video} exemplify this paradigm. Central to these methods is the solution of inverse problems that recover encoded object information from indirect measurements \cite{bertero2021introduction, tikhonov1963regularization, jin2017deep}. By integrating optical design with algorithmic inference, computational imaging expands the capabilities of traditional systems and enables enhanced performance across diverse applications, from biomedical imaging \cite{greenbaum2014wide, park2018quantitative, mait2018computational, eils2003computational} to remote sensing \cite{zhong2018computational, erkmen2012computational}.

\begin{figure}
    \centering
    \includegraphics[width=\linewidth]{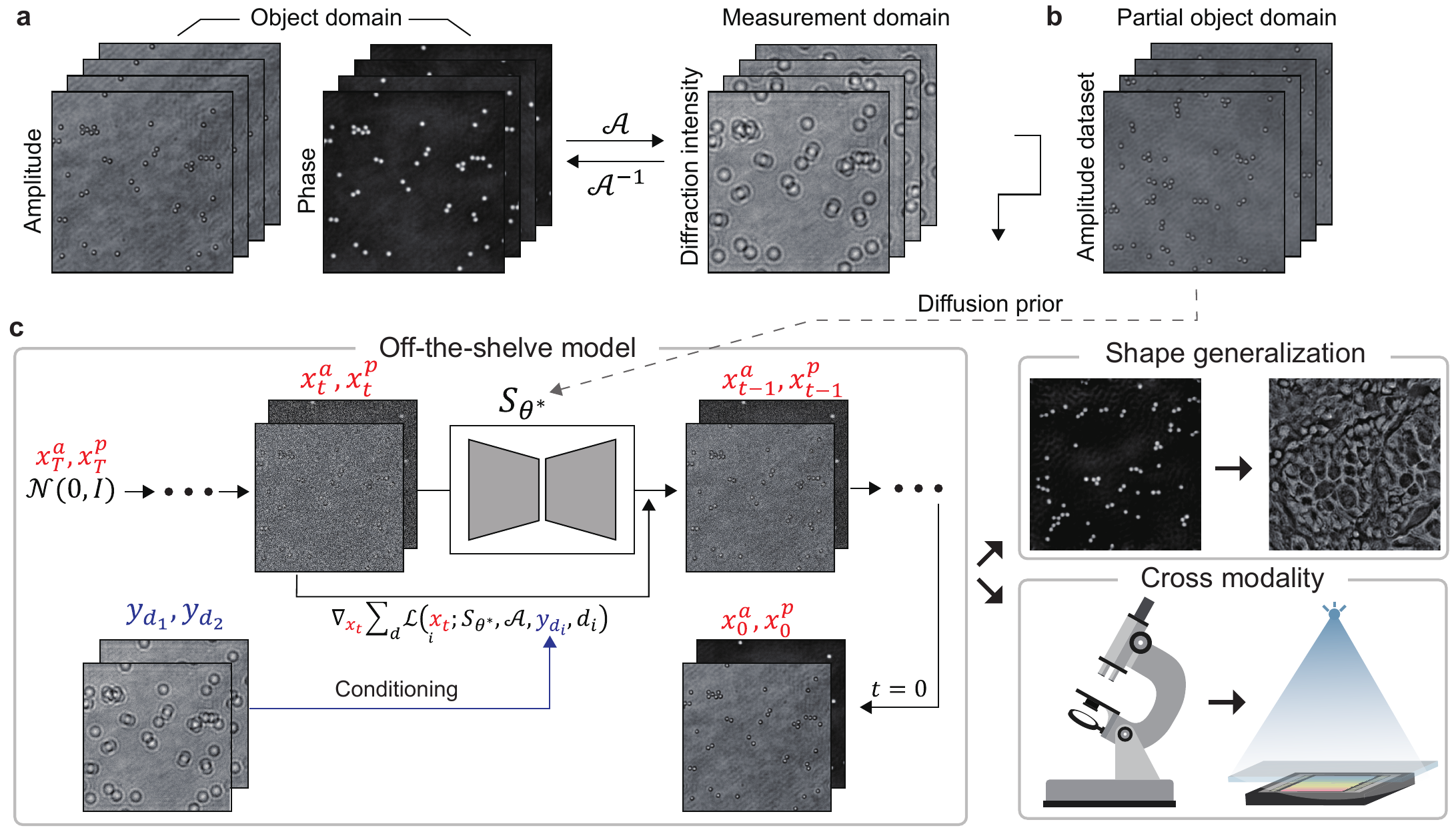}
    \caption{
    \textbf{Overall schemtic of the proposed method.}
    \textbf{a.} The object domain comprises both amplitude and phase components, whereas the measurement domain contains only diffraction intensity patterns.
    \textbf{b.} During training, low-dimensional amplitude images are used to learn a diffusion prior $S_{\theta^*}$, without requiring paired phase information.
    \textbf{c.} At inference time, the proposed method operates as an off-the-shelf model for complex amplitude reconstruction, demonstrating strong generalization across object shapes, imaging hardware, and even cross-modality scenarios.
    }
    \label{fig:overall}
\end{figure}

At the core of computational imaging lies the inverse problem, typically formulated as $y=\Ac(x)+\veps$, where $x$ is the true signal, $y$ is the measurement, $\Ac$ is the forward operator, and $\veps$ denotes stochastic noise such as Poisson shot noise. Many imaging systems involve nonlinear and many-to-one mappings, making the inverse problem ill-posed. Coherent imaging, in particular, exemplifies this challenge: sensors capture only light intensity, the squared magnitude of the complex field, resulting in loss of phase information. In-line holography \cite{schnars2014digital, garcia2006digital} is a representative case, where coherent light diffracts through a weakly scattering object and is recorded at the sensor as intensity measurements (Figure~\ref{fig:overall}a). The forward model, typically described via the optical diffraction model, involves nonlinear operations that couple amplitude and phase, requiring phase retrieval to recover the complex field:
\begin{equation}
    \Ib(x,y) = \mathcal{A}(\Ob(x, y); \phi) + \veps = \intensity{\transferfn{\Hb_\phi(x,y)} \ast \Ob(x,y)}+\veps,
    \label{eqn:angular_spectrum}
\end{equation}
where $\Ib(x,y)\in \mathbb{R}^{H\times W}$ and $\Ob(x,y)\in \mathbb{C}^{H\times W}$ denote the diffraction intensity (measurement) and the complex field (true signal), $\fourier$ and $\inv{\fourier}$ denote the matrices for 2D Discrete Fourier transform and its inverse transform,
and $\Hb_\phi$ denotes a transfer function parameterized by physical system properties $\phi$ (e.g., wavelength, sensor pixel size, object-to-sensor distance),

Recent advances in deep learning have shown promise in solving such nonlinear inverse problems \cite{wei2018deep, ongie2020deep, li2018deepnis, aggarwal2018modl, lee2023deep, huang2023self, barg2025adaptable, zeng2021deep, rivenson2019deep, rivenson2018phase, yu2023deep, wu2018extended, wang2019net, chen2023dh, ren2019end, chen2022fourier, liu2020deep, wang2020deep, qin2018convolutional, li2018deep, lee2025physics, kim2022deep}, but their effectiveness is often constrained by the need for large, annotated datasets and system-specific training. In holography, acquiring complex-valued ground truth (amplitude and phase) requires expensive interferometric setups, making dataset construction labor-intensive and impractical. Furthermore, deep learning models often fail to generalize across different imaging systems or object types. While recent works leveraging generative models such as CycleGANs \cite{lee2023deep} or training on simulation-based \cite{barg2025adaptable} or synthetic \cite{huang2023self} datasets have improved generalization to some extent, they remain sensitive to mismatches between training and deployment conditions. Alternative approaches that exploit architectural priors, such as Deep Image Prior \cite{wang2020phase} and Deep Decoder \cite{niknam2021holographic}, aim to bypass training but suffer from instability and limited robustness due to their reliance on implicit biases and manual tuning.

In this paper, we present an off-the-shelf solution for the phase retrieval problem in in-line holographic imaging. Our approach leverages a diffusion model~\cite{song2019generative, song2021scorebased, ho2020denoising} with posterior sampling, trained exclusively on amplitude data (Figure~\ref{fig:overall}b), yet capable of recovering both amplitude and phase information (Figure~\ref{fig:overall}c). This design allows us to address the inherent nonlinearity of the forward model using only partial, low-dimensional input, amplitude measurements that are more accessible and cost-effective in practical systems. The key contributions of our method are summarized as follows:

\begin{itemize}[label={$\bullet$}]
\item We introduce a diffusion model with posterior sampling trained solely on amplitude data to jointly recover amplitude and phase.
\item We develop a predictor-corrector framework with decoupled likelihood gradients for robust complex field reconstruction.
\item We demonstrate strong generalization to out-of-distribution (OOD) domains and system variability, even from benchtop to lensless on chip imaging configuration, in both simulations and experimental validations.
\end{itemize}

We validate the proposed framework under significantly more challenging conditions than typical OOD and hardware variability benchmarks (Figure~\ref{fig:overall}c). Specifically, a model trained soley on amplitude measurements of polystyrene beads, simple and repetitive in structure, is shown to reconstruct the complex amplitude of biologically derived samples that exhibit significantly higher structural and textural complexity. Remarkably, the posterior sampling trajectory adapts to unseen measurement conditions without retraining, enabling physically consistent phase retrieval across diverse scenarios. This adaptability holds across distinct acquisition geometries, including both benchtop holographic microscopes and lensless on-chip imaging systems, highlighting the method's robustness and practical utility. Overall, our framework demonstrates how learning from amplitude data, which can be readily acquired with conventional microscope setup, can effectively recover high-dimensional complex fields, offering a scalable and generalizable solution for a wide range of computational imaging problems.

\section{Results}\label{sec2}

\subsection{Bridging the gap: from amplitude-only data to complex amplitude reconstruction}
\label{subsec:recon}

We first evaluated our method’s object-field reconstruction capability using $3 \mu m$ polystyrene beads and tissue sections. To train the diffusion prior $s_{\theta}$, we acquired amplitude images of the polystyrene beads via a conventional microscopy setup. Unless otherwise specified, all experiments in this study employed a diffusion model trained exclusively on amplitude data from these  $3\mu m$ polystyrene beads. We obtained three diffraction intensities by displacing the sensor along the optical axis at distances 7, 10, and 15mm. Subsequently, we leveraged two intensities as conditions for the sampling of the pre-trained diffusion model, reconstructing the object field.

During inference, the trained diffusion model progressively denoises the amplitude $x_t^a$ and phase $x_t^p$, over a predetermined number of iterations $T$, where $t$ denotes the current iteration (Figure~\ref{fig:overall}c). This process is formulated as a stochastic differential equation (SDE) and solved numerically with a predictor-corrector (PC) framework~\citep{song2021scorebased}. After each "predictor" and "corrector" step, distinct likelihood gradients are applied to the amplitude and phase to enforce desired physical constraints.
Specifically, in the "predictor" step, we apply a variance-exploding (VE) sampling strategy~\citep{song2021scorebased} to denoise $x_t^a$ and phase $x_t^p$. Tweedie’s formula~\citep{efron2011tweedie, kim2021noise2score} is then used to obtain the posterior-mean amplitude $\hat{x}_0^a=x_t^a+\sigma_t^2\hat {s}$, where $\hat{s}=s_{\theta^*}(x_t, \sigma_t)$ denotes predicted score function by diffusion model, and $\sigma_t$ denotes a noise-related scaling factor~\citep{song2019generative}.
This operation offers average of final noise-free estimates of the amplitude as $t\rightarrow0$, avoiding reliance on noisy intermediate states for forward imaging computations (see \cref{eqn:hologrpahy_otf} in Methods section), and enabling effective enforcement of physical constraints. Following the predictor step, a "corrector" step updates the marginal distribution of the estimated sample $x_t \leftarrow x_t + \zeta_t \hat {s} + \sqrt{2\zeta_t}\epsilon$, where $\zeta_t$ is the step size and $\epsilon \sim N(0, \sigma^2)$. Physical constraints, again guided by Tweedie’s formula, were applied to both amplitude and phase to further refine the estimates, ultimately yielding $x_{t-1}^a$ and $x_{t-1}^p$~\citep{chung2022diffusion}.

Notably, our approach omits diffusion prior for $x_t^p$ from posterior sampling and relies primarily on the diffusion prior in the amplitude domain. 
In essence, the amplitude distribution is explicitly modeled by the diffusion prior, while the phase distribution is optimized through physical constraints—a strategy reminiscent of conventional iterative phase-retrieval algorithms such as Gerchberg–Saxton and Hybrid Input–Output. 
Because the distribution for $x_t^p$ is not explicitly learned, Tweedie’s formula is not applied to the phase variable; instead, its updates depend solely on measurement conditioning. Consequently, the diffusion prior drives $\hat{x}_t^a$ toward the learned amplitude distribution, whereas $x_t^p$ is iteratively refined to satisfy measurement constraints. Once the estimated amplitude distribution sufficiently converges with the trained prior, ultimately ensured by the sampling trajectory drifting toward the measured conditions, the proposed method achieves the desired complex amplitude reconstruction.

\begin{figure}
    \centering
    \includegraphics[width=\linewidth]{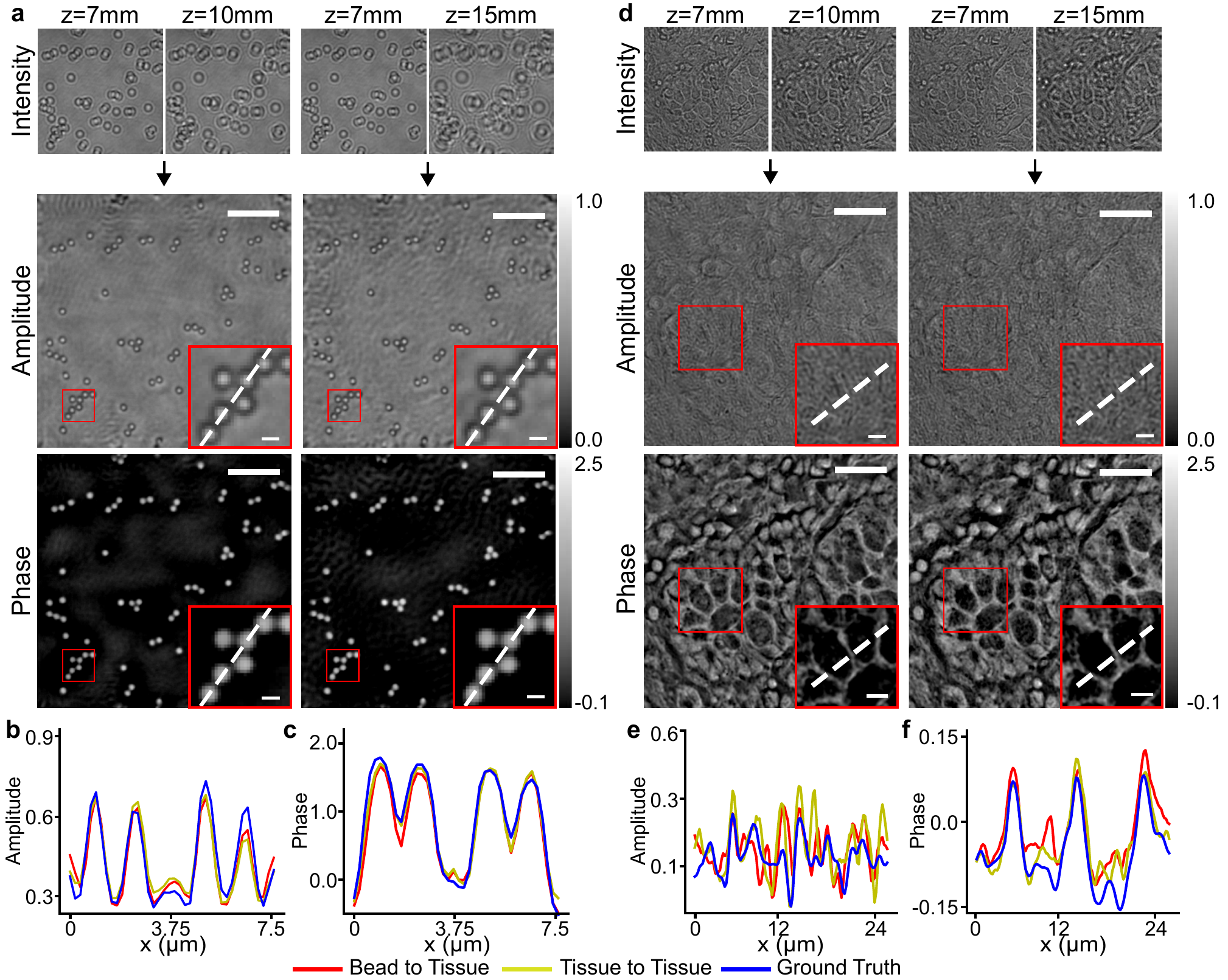}
    \caption{
    \textbf{Reconstructed complex amplitudes given diffraction intensities.}
    \textbf{a.} The proposed method effectively reconstructs the object field of polystyrene beads, regardless of the sample-to-sensor distance, using only two diffraction intensities. The scale bar in the full field of view indicates 10$\mu$m and the scale bar in insets indicates 1$\mu$m.
    \textbf{b, c.} The object amplitude and phase profiles along the white dotted lines on insets in a.
    \textbf{d.} The proposed method also works with tissue sections, even for the types that are not shown by the model during the training. The scale bar in the full field of view indicates 20$\mu$m and the scale bar in insets indicates 2$\mu$m.
    \textbf{e, f.} The object amplitude and phase profiles along the white dotted lines on insets in d.
    Red, Yellow: ours, Blue: Ground Truth.
    }
    \label{fig:exp2}
\end{figure}

As illustrated in Figure~\ref{fig:exp2}, the proposed method reconstructs the complex field of both beads and tissue sections using only two diffraction intensities as conditions (Figures~\ref{fig:exp2} a, d). To evaluate the precision of the reconstructed results, we computed the complex field profile and compared it with those of ground truth, which is experimentally measured from off-axis holography (Figure~\ref{fig:exp2} b,c,e,f).
The profiles show that the reconstructed complex fields are in good agreement with the ground truth, regardless of the sample-to-sensor distances of inline holography measurements. 
The significant advantage of the proposed method, namely its exceptional generalization capacity, is also demonstrated in these experimental results; for a more detailed analysis, see \cref{subsec:shape_general}.
Unlike conventional supervised learning schemes, the proposed method is not limited by a distance-dependent training process. This distinct feature enables the method to effectively reconstruct object fields for arbitrary sample-to-sensor distances. 

\begin{figure}[t]
    \centering
    \includegraphics[width=\linewidth]{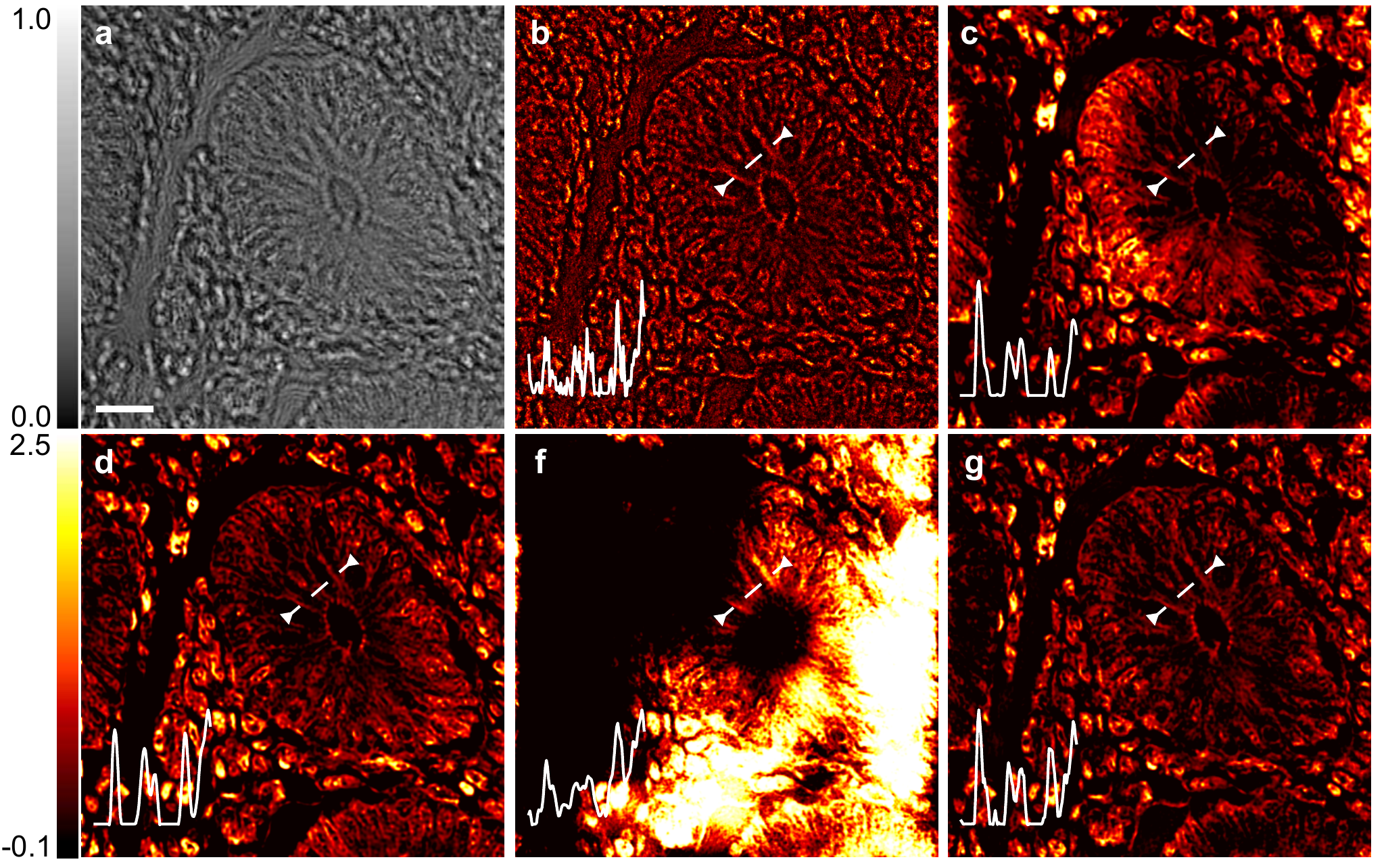}
    \caption{\textbf{Reconstruction comparison with the existing methods.}
    The target tissue section is the human appendix. The profile on the left-bottom side denotes the phase profile along a white dotted line on each field of view.
    \textbf{a.} Diffraction intensity measured at $15mm$. The scale bar indicates 20$\mu$m. \textbf{b.} Multi-height reconstruction. \textbf{c.} Deep Decoder reconstruction. \textbf{d.} Ground truth is measured by the off-axis holography. \textbf{f.} Deep Image Prior reconstruction. \textbf{g.} Proposed method reconstruction.}
    \label{fig:comp}
\end{figure}

We further evaluated the quality of complex field reconstruction by comparing our method with established holographic reconstruction techniques that, similar to our approach, do not require object phase information for training and rely solely on diffraction intensities during inference. Specifically, we selected three representative baselines: a projection-based multi-height phase retrieval algorithm~\cite{allen2001phase, greenbaum2012maskless}, and two neural network-based methods—Deep Decoder~\cite{niknam2021holographic} and Deep Image Prior~\cite{wang2020phase}. To ensure a fair comparison, all methods were provided with the same pair of diffraction intensity measurements for the inference stage.
As shown in Figure~\ref{fig:comp}, the proposed method provides better reconstruction results. While the multi-height algorithm successfully reconstructed the high-frequency structure of the tissue sample, the phase contrast between the tissue and background region is not correctly recovered. On the other hand, the Deep Decoder and Deep Image Prior generate a low-frequency artifact in their reconstructed phase while they adequately recover the contrast via different phase delays. Meanwhile, the proposed method reconstructs the object phase clearly and accurately.

\begin{figure}[t]
    \centering
    \includegraphics[width=\linewidth]{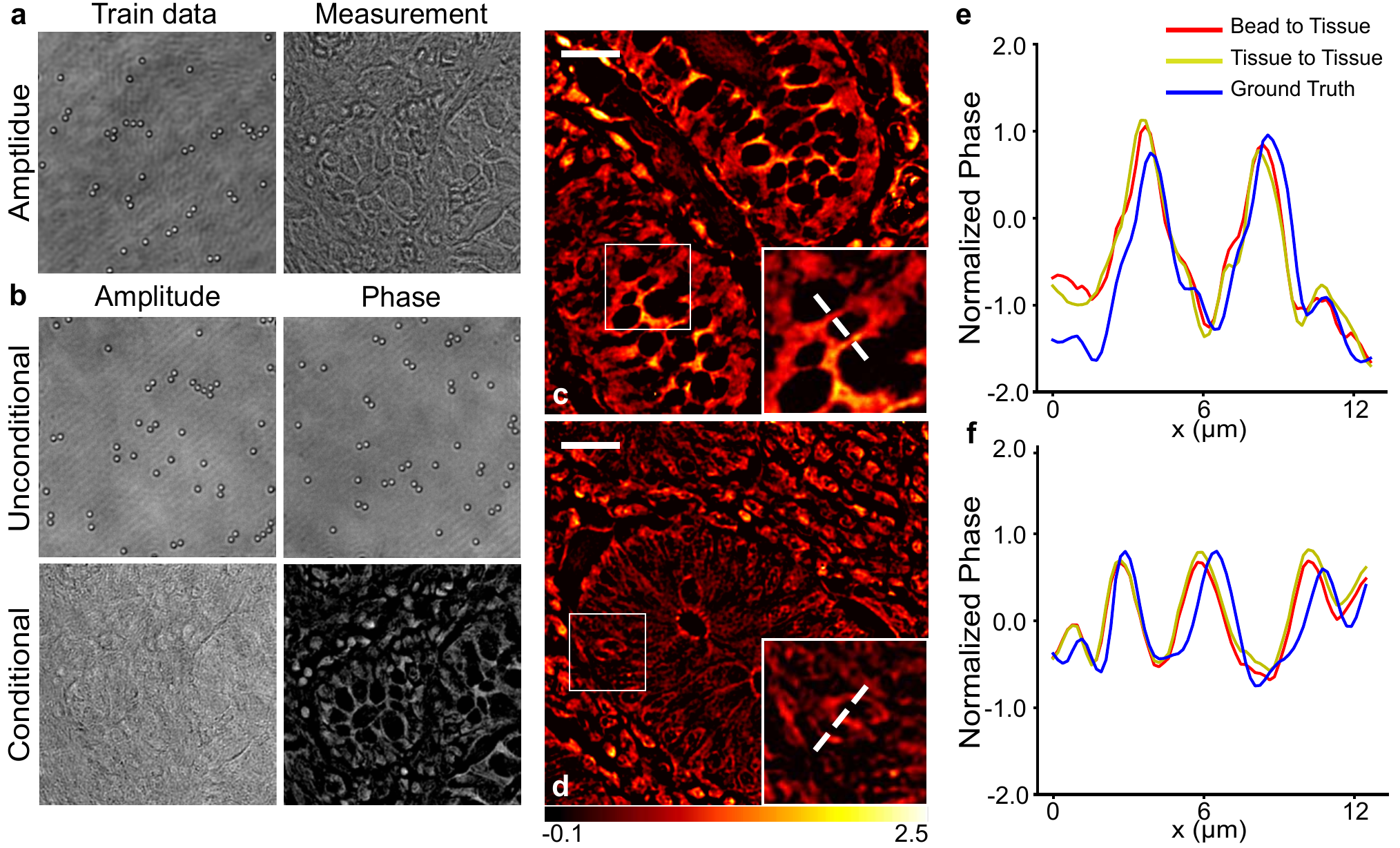}
    \caption{\textbf{Shape generalization capacity of the proposed method.}
    \textbf{a.} An example of polystyrene bead amplitude for the training and tissue section diffraction intensity for the inference.
    \textbf{b.} The proposed method trained on bead amplitudes generates random bead amplitude images for both amplitude and phase without conditions (the top row). However, the method generates tissue complex amplitude when conditions are given for the inference (the bottom row).
    \textbf{c, d.} Reconstructed phases of the small bowel and appendix by the proposed method trained with bead amplitude. The scale bar indicates 20$\mu$m.
    \textbf{e, f.} Phase profile along the white dotted line on c and d.The insets correspond to the white box in the full field of view.
    }
    \label{fig:shape_general}
\end{figure}

\subsection{Shape generalization of the proposed method}
\label{subsec:shape_general}

In practical applications involving deep learning models, a common approach has required substantial time to collect a sufficient amount of dataset and computational resources to train the neural networks before their deployment. Nevertheless, trained models frequently manifest constraints in their capacity to generalize, leading to a lack of universal applicability. In contrast, the proposed method provides significantly enhanced generalization capability that stems from test-time optimization. 
To validate the generalization capability of the proposed method for out-of-domain complex amplitude reconstruction, we applied the diffusion model trained on polystyrene bead amplitude for the reconstruction of complex fields of tissue section (Figure~\ref{fig:shape_general}).
%
The object amplitude used to train the diffusion model and the diffraction intensity used for inference are displayed in Figure~\ref{fig:shape_general}a. 
Without providing diffraction intensities as conditions, the trained diffusion model generates the amplitude of polystyrene beads for both amplitude and phase. However, if two diffraction intensities of the tissue section are given as conditions, the proposed method successfully restored the entire object field corresponding to the measurements (Figure~\ref{fig:shape_general}b).
To assess reconstruction accuracy, we compared the phase profiles from off-axis holography with those reconstructed by the diffusion model, showing strong agreement across various tissue types as illustrated in Figure~\ref{fig:shape_general}c–f. The experimental results highlight that the proposed method readily and effectively reconstructs object fields beyond its original domain, which suggests the potential for the method to serve as an off-the-shelf technique in the context of inline holography reconstruction.

\begin{figure}[t]
    \centering
    \includegraphics[width=\linewidth]{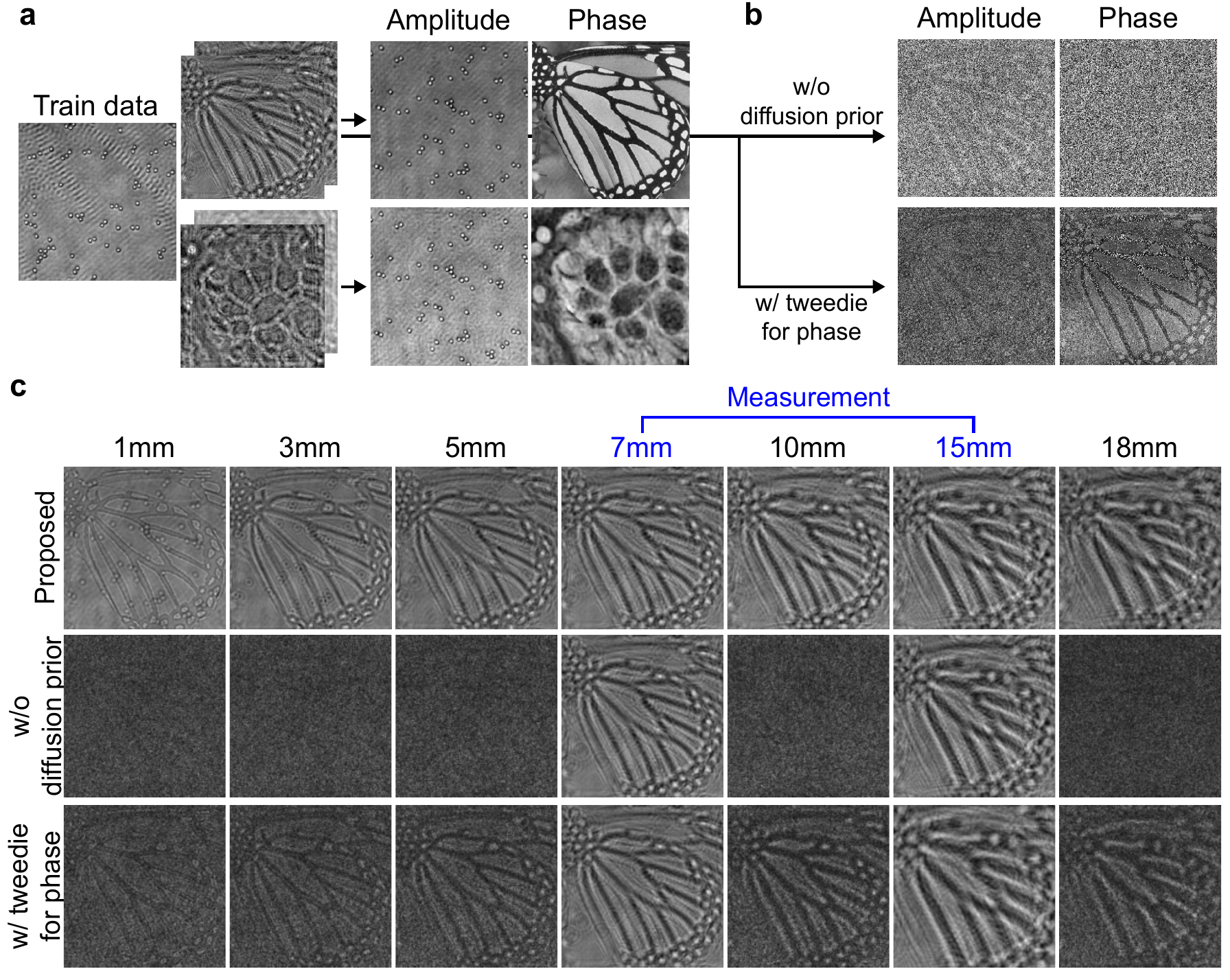}
    \caption{\textbf{Analaysis of shape generalization of the proposed method.}
    \textbf{a.} The proposed method reconstructs the object field where the diffraction intensities are synthesized by incorporating distinct amplitude and phase pairs.
    \textbf{b.} Ablation Study: Comparison of the reconstructed complex amplitude corresponding to the synthesized measurement. The reconstruction is shown for: (Top) A randomly initialized diffusion model (i.e., without a diffusion prior), and (Bottom) Applying the Tweedie's formula during the reconstruction procedure to estimate the clean phase image $\hat{x}_0^p$.
    \textbf{c.} Visualization of the sampling trajectory for the proposed method and the two ablation cases detailed in \textbf{b} (w/o diffusion prior and w/ tweedie for phase).
    }\label{fig:shape_general2}
\end{figure}

We further conducted an in-depth exploration of the proposed method through simulation studies. Specifically, we synthesized diffraction intensities using the angular spectrum method, with amplitude and phase maps intentionally decorrelated in morphology. In this process, we used the amplitude image of polystyrene beads in conjunction with two distinct phase counterparts: one derived from a tissue section and the other from a natural image. Remarkably, the proposed method demonstrates its exceptional shape generalization capability (by restoring complex amplitude accurately), not requiring the inherent similarity between amplitude and phase components (Figure~\ref{fig:shape_general2}a).
To discern whether the generalization performance predominantly stems from the conditioning methods, we conducted an ablation experiment in which the diffusion prior was omitted from the sampling procedure. 
The results reveal a deviation of the generated object field from the learned data manifold, leading to a notable reconstruction failure as shown in Figure~\ref{fig:shape_general2}b.
%
%

Additionally, when an inaccurate phase estimation is used by incorporating a diffusion model trained solely on amplitude data, similar reconstruction failures occur (Figure~\ref{fig:shape_general2}b).
Interestingly, in both cases, we observed that the reconstructions correspond to realizations of ill-posed solutions. In other words, the reconstructed field reproduces the measurement when the propagation distance matches that of the observation (Figure~\ref{fig:shape_general2}c). However, when the propagation distance is altered, the reconstructed complex field fails to yield a physically meaningful diffraction pattern. This behavior provides strong evidence of the problem's ill-posed nature and highlights the pivotal role of the diffusion prior. These findings demonstrate the proposed method's effectiveness by enabling superior shape generalization through test-time optimization within the context of posterior sampling.

The proposed method combines likelihood gradients with a predictor-corrector sampler, which serves as a numerical stochastic differential equation (SDE) solver. To illustrate how the likelihood gradients guide the sampling trajectory towards the conditioned distribution $p(\x|\y)$, we depict noisy amplitude and its clean estimation represented by the posterior mean $\mathbb{E}[\x_0|\x_t]$ in Figure~\ref{fig:shape_general2}c.
The sampling trajectory initially follows the learned prior informed by the training amplitude dataset. As sampling progresses, however, the likelihood gradients progressively steer the trajectory toward the measurement domain defined by the tissue data. This results in the reconstructed amplitude converging to match the observed human tissue measurements.
Meanwhile, since the diffusion model estimates the score function only for amplitude, the phase generation process is indirectly influenced through likelihood gradients defined on the noisy phase. Early in the sampling process, the phase is regulated by the evolving amplitude trajectory, as seen in the noisy phase reconstruction in Figure~\ref{fig:shape_general2}c. However, as $t\rightarrow 0$ and the amplitude becomes aligned with the measurement, the phase trajectory is similarly steered toward the measurement domain. This coupling arises from the data consistency update applied to the complex amplitude via the shared likelihood gradients.

%

\begin{figure}[t]
    \centering
    \includegraphics[width=\linewidth]{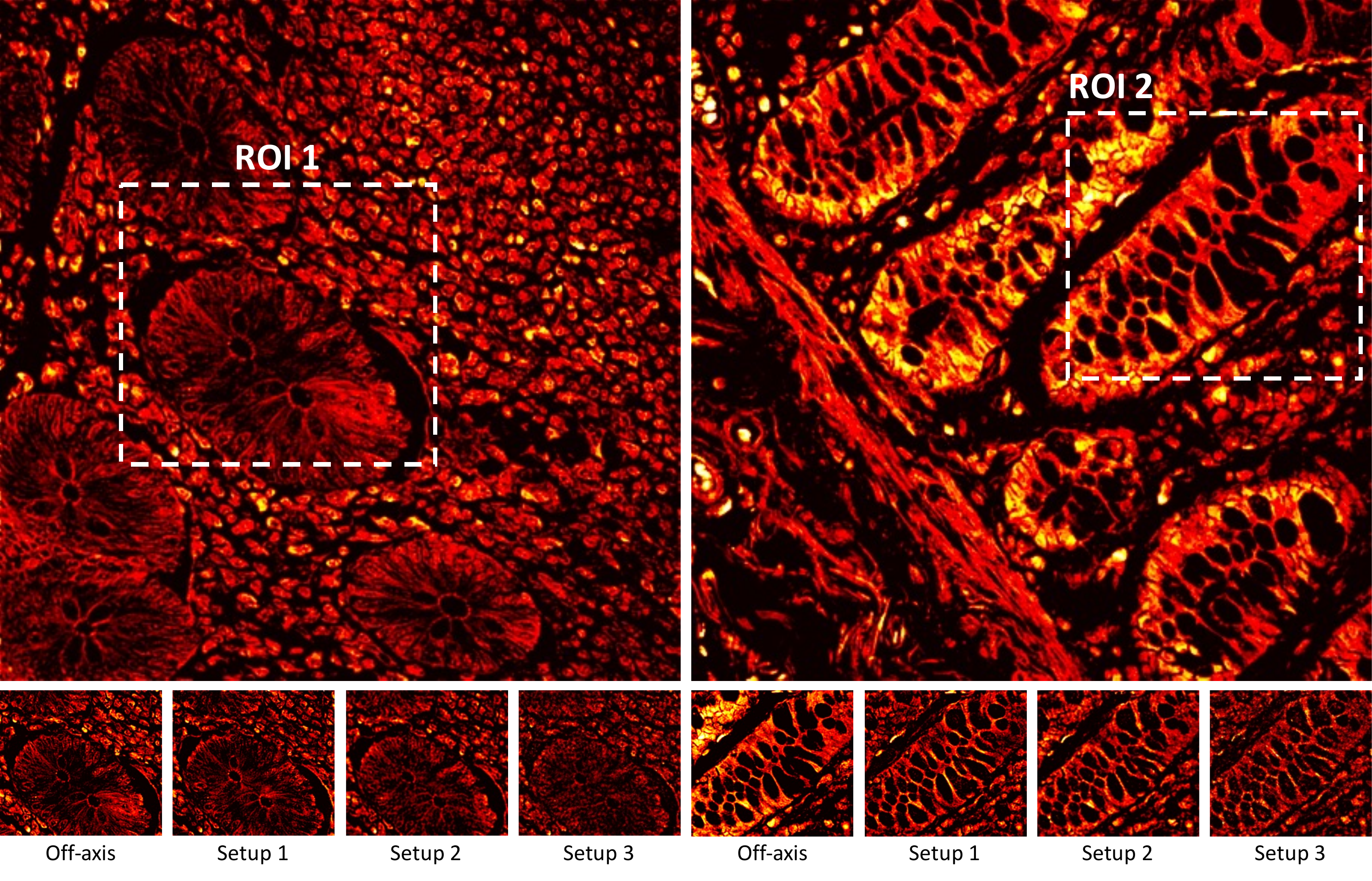}
    \caption{\textbf{Reconstructed object phase from measurements obtained with different effective pixel sizes.}
    The upper two images with larger FOV corresponds to object phase measured by the off-axis holography.
    Below views are the same ROI with white dotted rectangle in the upper view.
    Each setup used the same physical parameters except the magnification ratio and the sensor pixel size.
    Setup1: $300/9, 6.5\mu m$. Setup2: $300/18, 4.8\mu m$. Setup3: $300/18, 2.4\mu m$.
    }
    \label{fig:px}
\end{figure}

\subsection{Physical parameter generalization of the proposed method}
\label{subsec:param_general}
In clinical or field settings, holographic imaging systems often vary significantly depending on application-specific requirements, such as the size of the target object and the desired form-factor of the imaging device. As a result, imaging configurations can vary widely, and deep learning models trained on data from one setup typically do not generalize well to others. For example, in digital holographic microscopy, differences in the image sensor’s pixel size $\Delta p$ and system magnification $M$ alter the sampling rate of the object, while changes in the numerical aperture (NA) of the objective lens affect the maximum resolvable spatial frequencies. Additionally, the wavelength $\lambda$ of the illumination light influences light–matter interactions, causing the same biological sample to exhibit different optical responses under different illumination spectrum. As a result, the same specimen imaged under different system configurations can yield significantly different measurements, posing a key challenge for deep learning models trained under fixed conditions.

The generalization ability of the proposed method extends beyond data shape or domain, as it is inherently independent of the physical parameters of the imaging system during inference. To demonstrate this, we applied the proposed approach to diffraction intensity measurements acquired from a custom-built in-line holography system with varying imaging configurations. Specifically, we altered the magnification ratio of the 4f imaging setup to $300/18$ and used two different imaging sensors with pixel sizes of $4.8\mu m$ and $2.4\mu m$, respectively. We note that the training data composed of in-focus polystyrene beads acquired with a magnification factor of 300/9 and pixel size of $6.5\mu m$. From the standpoint of effective pixel size, defined as $\Delta p_{eff} = \Delta p/M$, these configurations correspond to downsampling and upsampling. As shown in Figure~\ref{fig:px}, the proposed method successfully reconstructs complex fields under all configurations. This result suggests that key components of the digital holographic imaging system, including the illumination source, optical components (e.g., objective lens), and image sensor, can be varied without compromising reconstruction performance. Such hardware adaptability highlights the practicality of the proposed method as an off-the-shelf holographic image solver and broadens its applicability across diverse system settings.


\begin{figure}[t!]
    \centering
    \includegraphics[width=\linewidth]{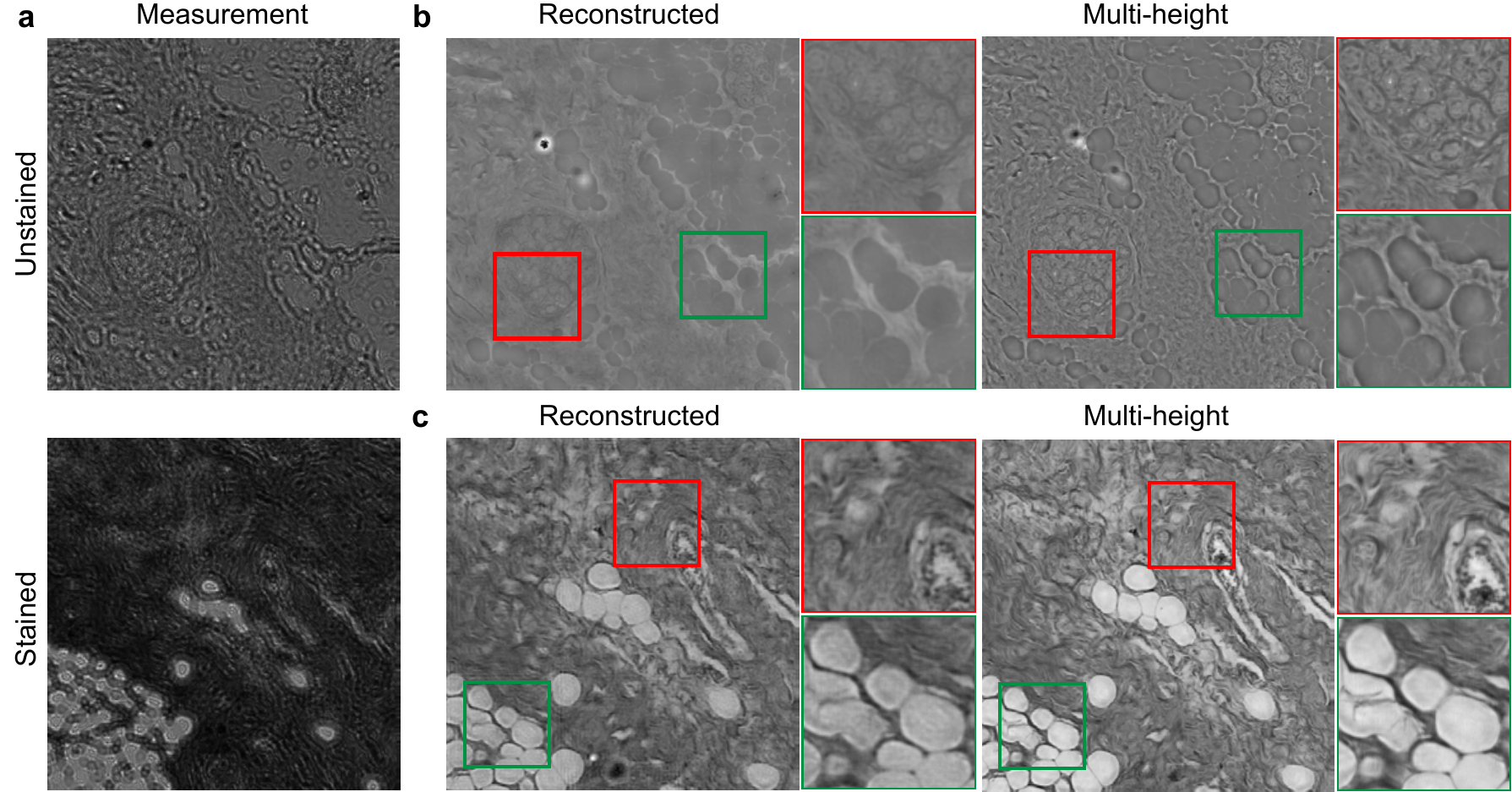}
    \caption{
    \textbf{Lensless imaging reconstruction results.}
    \textbf{a.} Examples of measurements obtained from the lensless system.
    \textbf{b.} Reconstructed object phase of unstained tissue section by the proposed method and multi-height algorithm.
    \textbf{c.} Reconstructed object amplitude of stained tissue section by the proposed method and multi-height algorithm.
    For the proposed method, two measurements are given as conditions, while six measurements are given for the multi-height algorithm.
    }
    \label{fig:lensless}
\end{figure}

\subsection{Cross-Modality Holography Using Diffusion Priors from Benchtop Microscopy in Lensless On-chip Tissue Imaging}
\label{subsec:lensless}

To demonstrate the broader cross-modal applicability of our approach, we evaluated its performance on lensless holographic imaging without any additional training data or modifications. Lensless on-chip imaging systems, which offer a simplified and compact form factor, are well-suited for clinical and biological applications such as in situ histopathology and real-time cell tracking. While the absence of an objective lens makes reconstruction more challenging due to the lack of band-limiting operation in spatial frequency, it also offers the advantage of a higher spatial-bandwidth product. Cross-modal transfer between benchtop microscopy and lensless on-chip imaging system is intrinsically challenging as they implement different acquisition geometries, magnifications, spatial-frequency band-limiting, and sampling regimes. To the best of our knowledge, zero-shot transfer from benchtop holography to lensless on-chip imaging—i.e., reconstructing on-chip measurements with a model trained only on benchtop data—has not previously been demonstrated.

To assess the practicality of our method as an off-the-shelf solution in the context of a lensless microscopy setup for clinical use, we acquired two diffraction intensity measurements of a breast tissue section at object-to-sensor distances of 0.8mm and 1.0mm using a simple lensless on-chip system composed of light source, specimen, and image sensor. lensless imaging setup, as illustrated in Figure~\ref{fig:lensless}a. These measurements were then used as input conditions for the sampling process of the diffusion model, which had been trained exclusively on in-focus amplitude images of polystyrene beads acquired with a benchtop microscope.

As shown in Figure~\ref{fig:lensless}b, c, the reconstructed amplitude and phase of each target have good agreement with the ground truth image.
Compared to benchtop microscope which is used for acquiring training samples, the lensless system is composed of completely different components such as an illumination source with different bandwidth, an image sensor with different characteristics (e.g. pixel size, dynamic range, and bit-dept), and also affected by different noise factors originate from measurement environment. Even in such extreme
case, the proposed method has demonstrated its adaptability and generalization capability.

\section{Discussion and Conclusion}
\label{conclusion}
In this work, we propose a diffusion-based method for complex amplitude reconstruction in holographic imaging, relying on only amplitude data for training. Unlike existing data-driven approaches that require ground-truth phase maps, the proposed scheme can be trained from amplitude-only images acquired on a standard benchtop microscope. This framework offers a ready-to-use, off-the-shelf solution that generalizes well across diverse object types and imaging configurations, even under significant shifts in the measurement domain (i.e., bead to tissue, removal of spatial band-limiting condition). The core contribution lies in demonstrating that low-dimensional amplitude information ($\mathbb{R}^N$) can serve as a strong constraint for recovering high-dimensional complex fields ($\mathbb{C}^N$, equivalently $\mathbb{R}^{2N}$) from nonlinearly coupled intensity measurements. Compared to prior diffusion-based methods, which focus on separate amplitude/phase reconstruction \cite{zhang2024single} or qualitative enhancement \cite{verma2025diffusion}, our approach enables joint complex amplitude recovery with robust generalization.

This strategy is broadly applicable to computational imaging modalities that aim to recover high-dimensional information from limited measurements. Applications such as coherent diffractive imaging, hyperspectral imaging, Fourier ptychographic microscopy, diffraction tomography, and single-pixel imaging all share a common challenge of jointly recovering high-dimensional signals from partial observations. In these settings, the proposed method can serve as an effective inverse solver when paired with a suitable forward model, offering a scalable and cost-efficient framework for reconstructing complex quantities from simple or undersampled data.

\section{Methods}
\label{sec:method}

\subsection{Sample preparation}
\label{subsec:sample_prepar}
$3\mu m$ polystyrene microspheres (PolyScience, refractive index 1.60) were diluted tenfold in a methanol solution and deposited onto a glass slide. After the methanol had evaporated, ultraviolet (UV) curing glue (Thorlabs, refractive index 1.54) was applied to the region covered by the microspheres to minimize the refractive-index mismatch between the embedding medium and the polystyrene beads. A coverslip was then placed on top to ensure uniform coverage, and UV irradiation was used to cure the adhesive, thereby stabilizing the sample. Tissue histopathology specimens were acquired from SuperBioChips Laboratories; all tissues presented in this study were drawn from Tissue Array (AA) and remained unstained.

\subsection{Experimental setup and data acquisition}
\label{subsec:method_exp}

We constructed custom benchtop microscope and a lensless microscopy configuration to generate the datasets used in this study. In bench top microscope, we employed a 532nm continuous-wave laser (Samba, Cobolt) as the illumination source. In the sample-beam path, a 4f imaging arrangement magnified the object, with an objective lens (20×, 0.25NA) for polystyrene bead samples and an objective lens (40×, 0.4NA) for histopathological tissues. Amplitude data in the object domain were acquired by capturing images at the focal plane of polystyrene beads, while diffraction intensity measurements in the measurement domain were collected from histopathological tissue samples. Additionally, ground-truth complex amplitude data, used solely for evaluation, were obtained through an interferometric setup.

In this becnh-top microscope, the measurement of the diffraction intensity can be represented as

\begin{equation}
    I(x,y) =  \left|\iint e^{-j(k_xx+k_yy)}OTF(k_x,k_y) \Tilde{H_\phi}(k_x,k_y) \Tilde{O}(k_x, k_y) dk_xdk_y \right|^2 +\epsilon
    \label{eqn:hologrpahy_otf}
\end{equation}
, where $\Tilde{O}$ and $\Tilde{H_\phi}$ denote the Fourier transforms of the object function and the imaging system’s transfer function, respectively, and $k_x$ and $k_y$ are spatial frequencies in the x and y directions. The term $OTF$ characterizes the amplitude and phase transfer functions of the objective lens. Due to the finite acceptance angle defined by the numerical aperture (NA), the transfer function of the objective lens behaves as a low-pass filter: $|OTF|=0$ for $|\bm{k}| >k_{NA}$ and $|OTF|=1$ otherwise. 

For lensless microscopy, a broadband LED source illuminates the sample placed directly above the image sensor, without any intervening lenses. Thus, the optical transfer function in the forward model \cref{eqn:hologrpahy_otf} is omitted, and the measured spatial frequency spectrum is instead determined by the temporal coherence of the illumination, determined by $\sqrt{z\Delta\lambda/8}$, where $\Delta\lambda$ denotes the spectral bandwidth of the source \cite{zhang2020resolution, you2024motion}. Also, under this partial coherence illumination provided by LED, higher angular components exceeding the coherent range appear as incoherent background fluctuation which can be considered as the noise term $\epsilon$ in \cref{eqn:hologrpahy_otf}. In the lensless microscopy setup, diffraction intensity measurements of histopathological tissue samples were captured at object-to-sensor distances of 0.8mm and 1.0mm. To obtain ground-truth complex amplitude data, additional intensity measurements were acquired at multiple axial positions, and a multi-height phase retrieval algorithm was applied for reconstruction.

\subsection{Implementation details}
We adopt the NCSN++ architecture proposed for VE-SDE~\cite{song2020score}, reducing the number of input and output channels to one. All experiments are carried out on square images of size $256\times256$, for which we both train the diffusion model and solve the phase-retrieval problem.
Training use the Adam optimizer~\citep{kingma2014adam} with a learning rate of $2\times10^{-4}$. For both the bead and tissue datasets, we  set a weight decay of 0, $\beta_1=0.9, \beta_2=0.999$, eps$=1e-8$, and reserve the first 5000 iterations for warm-up.
Following the score-matching formulation and time discretization for VE-SDE in \cite{song2019generative}, we employ 1000 timesteps during both training and sampling. The model is trained for 200 epochs, after which the loss converges. 
During sampling, we fix the step size $\eta$ of data consistency gradient to 0.5 for both bead and tissue diffusion models.
Because the trained model's spatial resolution is limited to $256\times256$, we can reconstruct only limited field of view directly, which is smaller than common requirements in biological imaging. To reconstruct larger field-of-view, we crop each in-line hologram into overlapping $256\times256$ patches with a 192-pixel stride. We then solve phase retrieval problem independently, crop the central $192\times192$ region of every reconstructions, and concatenate the results. This overlap-crop strategy yields seamless transition between patches while mitigating boundary artifacts introduced by the discrete Fourier transform used in the data-consistency gradient computation.

\subsection{Algorithm}

The proposed method is described in Algorithm~\ref{alg:pseudo}. For brevity, we omit the overlap-crop strategy and only provide the reconstruction algorithm for each patch. Our algorithm stems from Predictor-Corrector (PC) sampling method for Variance Exploding Stochastic Differential Equation (VESDE), proposed in \cite{song2021scorebased}.
After each predictor and corrector step, we refine both the amplitude and phase through a data-consistency update followed by an amplitude projection. 
Specifically, for the data consistency step, we first compute the sum of the mean-squared errors between the measured amplitudes and the simulated amplitudes generated by the forward $\Ac$ and complex field estimate that consists of the posterior mean $\hat\x_0^a$ and current phase $\x^t_p$. Taking the gradients of this loss with respect to the current amplitude $\x_t^a$ and the current phase $\x_t^p$~\citep{chung2022diffusion}, we adjust $\x_t^a$ and $\x_t^p$ via gradient descent method with a step size $\eta$.
Next, we replace the amplitude of the propagated field  $\Ac(\x_t, d_i)$ with the measured amplitude $y_{d_i}$ at each measurement distance $d_i$. This projection resembles the multi-height algorithm; however, our method differs in that the final complex field is obtained by averaging the projected samples from all distances. In our main experiments, we demonstrate that just two measured amplitudes at two distinct distances are sufficient to reconstruct the complete complex field.

\begin{figure}[t]
\begin{minipage}{.49\textwidth}
    \vspace{-0.7cm}
    \begin{algorithm}[H]
    \setstretch{1.4}
            \small
           \caption{Predictor at timestep $t$}
           \label{alg:predictor}
            \begin{algorithmic}[1]
             \Require $\x_t$, $\sigma_t$, $\sigma_{t-1}$, $\s_{\theta^*}$
             \State{$\veps \sim \Nc(0, \rmI)$}
             \State{$\hat\s \gets \s_{\theta^*}(\x_t, \sigma_t)$}
             \State{$\hat\x_0 \gets \x_t + \sigma^2_t \hat\s$}
             \State{$\x_{t-1} \gets \hat\x_0 - \sigma^2_{t-1} \hat\s + \sqrt{\sigma_t^2 - \sigma_{t-1}^2}\veps$}
             \State {\bfseries return} $\x_{t-1}, \hat\x_0$
            \end{algorithmic}
    \end{algorithm}
\end{minipage}
\begin{minipage}{.49\textwidth}
    \vspace{-0.7cm}
    \begin{algorithm}[H]
            \small
           \caption{Corrector at timestep $t$}
           \label{alg:corrector}
            \begin{algorithmic}[1]
             \Require $\x_t$, $\sigma_t$, $\s_{\theta^*}$
              \For{$i=1$ {\bfseries to} $M$}
                 \State{$\veps \sim \Nc(0, \rmI)$}
                 \State{$\hat\s \gets \s_{\theta^*}(\x_t, \sigma_t)$}
                 \State{$\x'_t \gets \x_t + \zeta_t \hat\s + \sqrt{2\zeta_t}\veps$}
              \EndFor
              \State{$\hat\x_0 \gets \x_t + \sigma^2_t \s_{\theta^*}(\x_t, \sigma_t)$}
              \State {\bfseries return} $\x_t, \hat\x_0$
            \end{algorithmic}
    \end{algorithm}
\end{minipage}
\vspace{-0.5em}
\end{figure}

\begin{algorithm}[t]
\small
\caption{PC sampling (VPSDE) with data consistency with multiple measurements}
\label{alg:pseudo}
\begin{algorithmic}[1]
    \Require Number of timesteps $T$, Pre-defined coefficient $\sigma_t$, Number of measurements $n$, Set of measurement distances $d=\{d_1, ..., d_n\}$, Set of measured amplitudes via inline holography $y_d=\{y_{d_1}, ..., y_{d_n}\}$, Data consistency gradient step size $\eta$, Forward operator $\mathcal{A}$.
    
    \State{$\x^{a}_T, \x^{p}_T \sim \Nc(0, \rmI)$}
    \For{$t=T$ {\bfseries to} $1$}
        \State{$\veps^{a}, \veps^{p} \sim \Nc(0, \rmI)$}
        \State{$\x_t^{a},\hat\x_0^{a} \gets \text{Predictor}(\x_t^{a}, \sigma_t, \sigma_{t-1})$} \Comment{Amplitude predict}
        \State{$\x_t^{p}, \_ \gets \text{Predictor}(\x_t^{p}, \sigma_t, \sigma_{t-1})$} \Comment{Phase predict}
        \State{\color{purple}$\x_t^{a}, \x_t^{p} \gets \text{Data Consistency}(\hat\x_t^a, \x_t^p, d, y_d)$}\Comment{Algorithm \ref{alg:dc}}
        \State{\color{purple}$\x_t^{a}, \x_t^{p} \gets \text{Projection}(\x_t^{a}, \x_t^{p}, d, y_d)$}\Comment{Algorithm \ref{alg:proj}}
        \State{$\hat\x_t^{a},\hat\x_0^{a} \gets \text{Corrector}(\x_t^{a}, \sigma_t, \sigma_{t-1})$} \Comment{Amplitude correction}
        \State{$\hat\x_t^{p},\_ \gets \text{Corrector}(\x_t^{p}, \sigma_t, \sigma_{t-1})$} \Comment{Phase correction}
        \State{\color{purple}$\x_t^{a}, \x_t^{p} \gets \text{Data Consistency}(\hat\x_t^a, \x_t^p, d, y_d)$}\Comment{Algorithm \ref{alg:dc}}
        \State{\color{purple}$\x_t^{a}, \x_t^{p} \gets \text{Projection}(\x_t^{a}, \x_t^{p}, d, y_d)$}\Comment{Algorithm \ref{alg:proj}}
    \EndFor
    \State {\bfseries return} $\x_t^{a}, \x_t^{p}$
\end{algorithmic}
\end{algorithm}

\begin{figure}[t]
\begin{minipage}{.49\textwidth}
    \vspace{-0.7cm}
    \begin{algorithm}[H]
    \setstretch{1.4}
            \small
           \caption{Data Consistency}
           \label{alg:dc}
            \begin{algorithmic}[1]
             \Require $\hat\x_t^a$, $\x_t^p$, $\eta$ $d=\{d_1,...,d_n\}$, $y_d=\{y_{d_1},...,y_{d_n}\}$
             \State{$\hat\x_t^p \gets \pi(2\hat\x_t^p - 1)$} \Comment{Rescale to (-$\pi$, $\pi$)}
             \State{$\hat\x \gets |\hat\x_0^{a}|\cdot exp (1j\cdot\x_t^{p})$} \Comment{Complex field}
             \State{$\x_t^{a} \gets \x_t^{a} - \eta \nabla_{\x_t^{a}} \sum_{d_i}\|\y_{d_i}-|\Ac(\hat\x, d_i)|\|^2$}
             \State{$\x_t^{p} \gets \x_t^{p} - \eta \nabla_{\x_t^{p}} \sum_{d_i} \|\y_{d_i}-|\Ac(\hat\x, d_i)|\|^2$}
             \State {\bfseries return} $\x_t^{a}, \x_t^{p}$
            \end{algorithmic}
    \end{algorithm}
\end{minipage}
\begin{minipage}{.49\textwidth}
    \vspace{-0.7cm}
    \begin{algorithm}[H]
            \small
           \caption{Projection}
           \label{alg:proj}
            \begin{algorithmic}[1]
             \Require $\hat\x_t^a$, $\x_t^p$, $d=\{d_1,...,d_n\}$, $y_d=\{y_{d_1},...,y_{d_n}\}$
             \State{$\hat\x_t^p \gets \pi(2\hat\x_t^p - 1)$}\Comment{Rescale to (-$\pi$, $\pi$)}
             \State{$\hat\x \gets |\hat\x_t^{a}|\cdot exp (1j\cdot\x_t^{p})$} \Comment{Complex field}
              \For{$i=1$ {\bfseries to} $n$}
                 \State{$\hat\x_i \gets \mathcal{A}\left(\frac{\mathcal{A}(\hat\x_i, d_i)}{|\mathcal{A}(\hat\x_i, d_i)|} y_{d_i},-d_i\right) $}
              \EndFor
              \State{$\hat\x \gets \sum_i\hat\x_i /n$}
              \State {\bfseries return} $|\hat\x|, \angle \hat\x$
            \end{algorithmic}
    \end{algorithm}
\end{minipage}
\vspace{-0.5em}
\end{figure}

\subsection{Compared methods}
To evaluate the performance of the proposed method, we compare it with three other reconstruction techniques: the multi-height phase retrieval algorithm, the physics-enhanced deep neural network (PhysenNet)\cite{wang2020phase}, and the Deep Compressed Object Decoder (DCOD)\cite{niknam2021holographic}. The multi-height phase retrieval algorithm is a conventional holographic reconstruction approach that uses multiple diffraction intensities as amplitude constraints, while iteratively refining the phase. In contrast, PhysenNet and DCOD are iterative methods that leverage deep image priors (PhysenNet) or deep decoders (DCOD) as inherent priors for generating natural images, integrating these priors with the forward imaging model to solve for the unknown complex field. The original implementations of \href{https://github.com/FeiWang0824/PhysenNet}{PhysenNet} and \href{https://github.com/farhadnkm/DCOD}{DCOD} were adopted from their respective publicly available repositories, ensuring consistency with the methods as described by the authors.






\clearpage
\bibliographystyle{sn-nature}
\bibliography{sn-bibliography}


\end{document}

%% file: packages.tex
\usepackage{graphicx}%
\usepackage{multirow}%
\usepackage{amsmath,amssymb,amsfonts}%
\usepackage{amsthm}%
\usepackage{mathrsfs}%
\usepackage[title]{appendix}%
\usepackage{textcomp}%
\usepackage{manyfoot}%
\usepackage{booktabs}%
\usepackage{algorithm}%
\usepackage{algorithmicx}%
\usepackage{algpseudocode}%
\usepackage{listings}%

\usepackage[utf8]{inputenc} 
\usepackage[T1]{fontenc}    
\usepackage{hyperref}       
\usepackage{url}            
\usepackage{booktabs}       
\usepackage{nicefrac}       
\usepackage{microtype}      
\usepackage[table, dvipsnames]{xcolor}
\usepackage{soul}
\usepackage{times}
\usepackage{epsfig}
\usepackage{placeins}
\usepackage[skip=5pt]{caption}
\usepackage{rotating}
\usepackage{cancel}
\usepackage{setspace}
\usepackage{eso-pic}
\usepackage{tabu}
\usepackage{cellspace}
\usepackage{xkcdcolors}

\usepackage{bm}
\usepackage{bibunits} 

\usepackage{hyperref}
\usepackage{amsmath}
\usepackage{graphicx}
\usepackage{wrapfig,lipsum}

\usepackage{epsfig}
\usepackage{tikz}
\usetikzlibrary{spy}
\usepackage{algpseudocode}
\usepackage{algorithm}
\usepackage{mathrsfs}

\usepackage{nicefrac}       
\usepackage{booktabs}       

\usepackage{thmtools,thm-restate}
\usepackage{cleveref}

\usepackage{subcaption}        

%% file: macros.tex
\definecolor{mylightblue}{rgb}{0.85, 0.90, 0.94}
\sethlcolor{mylightblue}
\definecolor{mylightred}{rgb}{0.90, 0.85, 0.94}

\newcommand{\x}{{\boldsymbol x}}
\newcommand{\s}{{\boldsymbol s}}
\newcommand{\y}{{\boldsymbol y}}

\newcommand{\Ib}{{\boldsymbol I}}
\newcommand{\Ob}{{\boldsymbol O}}

\newcommand{\Hb}{{\boldsymbol H}}

\newcommand{\Nc}{{\mathcal N}}

\newcommand{\Ac}{{\mathcal A}}

\newcommand{\veps}{\boldsymbol{\epsilon}}
\newcommand{\rmI}{\boldsymbol{I}}
\def\eqref#1{(\ref{#1})}

\newcommand{\transferfn}[1]{\mathcal{F}^{-1}{#1}\mathcal{F}}
\newcommand{\intensity}[1]{\lvert{#1}\rvert^2}
\newcommand{\fourier}{\mathcal{F}}
\newcommand{\inv}[1]{{#1}^{-1}}

\definecolor{trolleygrey}{rgb}{0.5, 0.5, 0.5}
\definecolor{darkpastelgreen}{rgb}{0.01, 0.75, 0.24}
\definecolor{darkpink}{rgb}{0.91, 0.33, 0.5}
\definecolor{alizarin}{rgb}{0.82, 0.1, 0.26}
\definecolor{americanrose}{rgb}{1.0, 0.01, 0.24}

\makeatletter
\newcommand\footnoteref[1]{\protected@xdef\@thefnmark{\ref{#1}}\@footnotemark}
\makeatother